\documentclass{ws-procs9x6}

\begin{document}

\title{Nonperturbative Flipped $SU(5)$ Vacua in Ho\v rava--Witten
Theory}

\author{Alon E. Faraggi, Richard Garavuso}

\address{Department of Theoretical Physics,\\ 
1 Keble Road, Oxford OX1 3NP, UK\\ 
E-mail:  faraggi@thphys.ox.ac.uk, garavuso@thphys.ox.ac.uk}

\author{Jos\'e M. Isidro}

\address{Instituto de F\'{\i}sica Corpuscular (CSIC--UVEG)\\
Apartado de Correos 22085, 46071 Valencia, Spain\\
E-mail: jmisidro@ific.uv.es}  


\maketitle

\abstracts{There is good  support for the embedding of
the Standard Model fermions in the chiral ${\bf 16}$ $SO(10)$ representation.
Such an embedding is provided  by the realistic free fermionic
heterotic--string models. In this talk we demonstrate the
existence of solutions  with three generations and
$SO(10)$ observable gauge group, in the case of compactification on 
an torus--fibred Calabi--Yau space over a  Hirzebruch base surface.
The $SO(10)$ symmetry is broken to $SU(5)\times U(1)$
by a Wilson line. The overlap with the realistic free fermionic
heterotic--string models is discussed.}

Invited talk presented by J. M. I. at the 1st International
Conference on String Phenomenology, Oxford, 6-11 July 2002.

\section{Introduction}\label{suru}

There has been considerable interest in the construction
realistic particle physics vacua with $N=1$ supersymmetry, three families
of quarks and leptons and a grand unified gauge group $H$ upon 
compactification 
of Ho\v rava--Witten theory \cite{hw} on a complex Calabi--Yau 3--fold 
\cite{dlow}. The latter can be an elliptically--fibred manifold $X$ over a 
base complex 2--fold $B$ or, more generally, it can be a torus--fibred 
3--fold $Z$. While $X$ is assumed to possess a section 
$\sigma:B\rightarrow X$, 
$Z$ need not carry a section. 
The $E_8$ gauge group on the observable sector decomposes 
as $E_8\supset G\times H$, where $H$ is the gauge group of the effective 
field theory and $G$ appears as the structure group of a holomorphic, 
stable vector bundle over the 3--fold. We are interested in the case when
$G=SU(4)$ and $H=SO(10)$.
A nonperturbative vacuum state of the resulting GUT theory 
on the observable sector is specified by a set of 
M--theory 5--branes wrapping a holomorphic 2--cycle on the 3--fold. The 
5--branes are described by a 4--form cohomology class $[W]$
satisfying the anomaly--cancellation condition. 
This class is Poincar\'e--dual 
to an effective cohomology class in $H_2(X, {\bf Z})$ that can be written as
\begin{equation}
[W]=\sigma_*(\omega) + c(F-N) + dN,
\label{eq:clascur}
\end{equation}
where $c,d$ are integers, $\omega$ is a class in $B$, and
$\sigma_*(\omega)$ is its pushforward to $X$ under $\sigma$.
The rules to construct these vacua explicitly can be summarised
as follows (see refs. \cite{dlow},\cite{noi} for more details):

a) Semistability condition: the spectral data specifying a
semistable, holomorphic vector bundle on the 3--fold can be written in terms 
of an effective class $\eta\in H^2(B, {\bf Z})$ 
and coefficients $\lambda$, $\kappa_i$ 
satisfying
\begin{equation}
\lambda\in {\bf Z}, \qquad \eta=c_1(B)\;{\rm mod}\; 2,
\label{eq:condlamx}
\end{equation}
with the $\kappa_i$ either all integer or all half an odd integer,
or alternatively
\begin{equation}
\lambda= {2m+1\over 2}, \, m\in {\bf Z}, \qquad c_1(B) \;
{\rm even},
\label{eq:altsetxz}
\end{equation}
with the same requirements on the $\kappa_i$. Above, $c_1$ denotes the 
first Chern class.

b) Involution conditions: for a vector bundle $V_X$ on $X$ to
descend to a vector bundle $V_Z$ on $Z$ it is necessary that
\begin{equation}
\tau_B(\eta)=\eta,\qquad \sum_i\kappa_i=\eta\cdot c_1(B).
\label{eq:invcondxz}
\end{equation}

c) Effectiveness condition: a sufficient condition for $[W]$ in eqn.
(\ref{eq:clascur}) to be an effective class is
\begin{equation}
12c_1(B)\geq \eta,\qquad c\geq 0,\qquad d\geq 0.
\label{eq:efectdonc}
\end{equation}

d) Commutant condition: for $H=SO(10)$ this condition reads \cite{bmr}
\begin{equation}
\eta\geq 4 c_1(B).
\label{eq:bergx}
\end{equation}

e) Three--family condition:
\begin{equation}
\lambda\eta\left(\eta - n c_1(B)\right)=6.
\label{eq:nogenxz}
\end{equation}

\section{Vacua over Hirzebruch surfaces $F_r$}\label{hirsur}

We take the base manifold $B$ to be the Hirzebruch surface $F_r$,
$r\geq 0$.
The latter is a ${\bf CP}^1$--fibration over ${\bf CP}^1$.
A basis for $H_2(F_r, {\bf Z})$ composed of effective classes
is given by the class of the base ${\bf CP}^1$,
denoted $S$, plus the class of the fibre ${\bf CP}^1$,
denoted $E$. Their intersections are $
S\cdot S= -r$, $S\cdot E=1$,
$E\cdot E=0$.
All effective curves in $F_r$ are linear combinations of $S$ and
$E$ with nonnegative coefficients. The Chern classes of $F_r$ are
$c_1(F_r)=2S + (r+2) E$, $ c_2(F_r)=4$.
It is proved in ref. \cite{dlow} that, over the base $F_r$, one can construct
torus--fibred Calabi--Yau 3--folds $Z$ whose fundamental group is ${\bf Z}_2$
when $r=0,2$. For those allowed values of $r$, any class $\eta\in H_2(F_r, 
{\bf Z})$ is $\tau_B$--invariant \cite{dlow}.
In what follows we will work with an arbitrary allowed value of $r$.
Let us write $\eta\in H_2(F_r, {\bf Z})$ as
$
\eta=sS + e E$,
for some integers $s, e$ to be determined imposing the conditions summarized
in section \ref{suru}.
We can now go to eqn. (\ref{eq:clascur}) and write explicit expressions for
the homology class $[W]$ that is being wrapped by the fivebranes on the
torus--fibred Calabi--Yau 3--fold $Z$.  We have 
$\omega=(24-s)S + (12r+24-e)E$,
\begin{equation}
c=112+{3\over \lambda} -12\lambda -\sum_i\kappa_i^2,
\label{eq:ece}
\end{equation}
\begin{equation}
d=16+{3\over \lambda} - 12\lambda +\sum_i\kappa_i -\sum_i\kappa_i^2.
\label{eq:ede}
\end{equation}
Every allowed choice of $r$, plus every
choice of the rational coefficients $\kappa_i$ subject to the conditions
indicated
in each case, gives rise to a different vacuum $[W]$:

$\bullet$ $s=9$: $\sum_i\kappa_i=30$ and $\sum_i\kappa_i^2\leq 46$,
\begin{equation}
[W]=\sigma_*\left(15S+\left({15\over 2}r + 18\right) E\right)+
(112-\sum_i\kappa_i^2)(F-N) +
(46-\sum_i\kappa_i^2)N.
\label{eq:snove}
\end{equation}

$\bullet$ $s=10$: $\sum_i\kappa_i=34$ and $\sum_i\kappa_i^2\leq 34$,
\begin{equation}
[W]=\sigma_*\left(14 S + (7r + 17) E\right) +
(96-\sum_i\kappa_i^2) (F-N) +
(34 -\sum_i\kappa_i^2)N.
\label{eq:sdiec}
\end{equation}

$\bullet$ $s=11$: $\sum_i\kappa_i=34$ and $\sum_i\kappa_i^2\leq 66$,
\begin{equation}
[W]=\sigma_*\left(13S+\left({13\over 2} r + 18\right)E\right) +
(128 - \sum_i \kappa_i^2)(F-N) +
(66 - \sum_i\kappa_i^2)N.
\label{eq:sundici}
\end{equation}

$\bullet$ $s=13$:  $\sum_i\kappa_i=38$ and $\sum_i\kappa_i^2\leq 38$,
\begin{equation}
[W]=\sigma_*\left(11 S + \left({11\over 2} r + 18 \right) E\right) +
(96 - \sum_i\kappa_i^2)(F-N) +
(38 - \sum_i\kappa_i^2)N.
\label{eq:sdodici}
\end{equation}

$\bullet$ $s=14$: $\sum_i\kappa_i=38$ and $\sum_i\kappa_i^2\leq 54$,
\begin{equation}
[W]=\sigma_*\left(10S + (5r + 19)E\right) +
(112 - \sum_i\kappa_i^2)(F-N) +
(54 -\sum_i\kappa_i^2)N.
\label{eq:squatto}
\end{equation}

$\bullet$ $s=15$: $\sum_i\kappa_i=42$ and $\sum_i\kappa_i^2\leq 58$,
\begin{equation}
[W]=\sigma_*\left(9S + \left({9\over 2}r+18\right)E\right) +
(112-\sum_i\kappa_i^2)(F-N) +
(58-\sum_i\kappa_i^2)N.
\label{eq:squindici}
\end{equation}

$\bullet$ $s=18$: $\sum_i\kappa_i=46$ and $\sum_i\kappa_i^2\leq 78$,
\begin{equation}
[W]=\sigma_*\left(6 S + (3r+19)E\right) +
(128-\sum_i\kappa_i^2)(F-N) +
(78-\sum_i\kappa_i^2)N.
\label{eq:sdiciotto}
\end{equation}

$\bullet$ $s=22$: $\sum_i\kappa_i=54$ and $\sum_i\kappa_i^2\leq 54$,
\begin{equation}
[W]=\sigma_*\left(2S + (r+19)E\right) +
(96-\sum_i\kappa_i^2) (F-N) +
(54-\sum_i\kappa_i^2)N.
\label{eq:sventidue}
\end{equation}

\section{Overlap with the free fermionic models}\label{overlap}

In this section we elaborate briefly on the overlap with the free fermionic
models. Amazingly enough, the structure of the manifolds constructed in 
ref.  \cite{dlow}, up to the imposition of the three generation
condition, precisely coincides with the structure of the
manifold that underlies the free fermionic models.

In the free fermionic formalism \cite{fff} a model is specified in terms
of a set of boundary condition basis vectors and one--loop
GSO projection coefficients. These fully determine the
partition function of the models, the spectrum and
the vacuum structure. The three generation models
of interest here are constructed in two stages. The first
corresponds to the NAHE set of boundary basis vectors
$\{{\bf1},S,b_1,b_2,b_3\}$ \cite{nahe}. The second consists
of adding to the NAHE set three additional boundary
condition basis vectors, typically denoted $\{\alpha,\beta,\gamma\}$.
The gauge group at the level of the NAHE set is $SO(6)^3\times
SO(10)\times E_8$, which is broken to $SO(4)^3\times U(1)^3\times
SO(10)\times SO(16)$ by the vector $2\gamma$. Alternatively,
we can start with an extended NAHE set $\{{\bf1},S,\xi_1,\xi_2,b_1,
b_2\}$, with $\xi_1={\bf1}+b_1+b_2+b_3$. The set $\{{\bf1},S,\xi_1,
\xi_2\}$ produces a toroidal Narain model with $SO(12)\times
E_8\times E_8$ or $SO(12)\times SO(16)\times SO(16)$ gauge
group depending on the GSO phase $c({\xi_1\atop\xi_2})$.
The basis vectors $b_1$ and $b_2$ then break $SO(12)\rightarrow
SO(4)^3$, and either $E_8\times E_8\rightarrow E_6\times U(1)^2\times E_8$
or $SO(16)\times SO(16)\rightarrow SO(10)\times U(1)^3\times SO(16)$.
The vectors $b_1$ and $b_2$ correspond to ${\bf Z}_2\times {\bf Z}_2$ orbifold
modding. The three sectors $b_1$, $b_2$ and $b_3$ correspond to
the three twisted sector of the ${\bf Z}_2\times {\bf Z}_2$ orbifold,
with each producing eight generations in the ${\bf27}$, or
${\bf16}$, representations
of $E_6$, or $SO(10)$, respectively. In the case of $E_6$ the untwisted
sector produces an additional $3\times ({\bf 27}+{\overline{\bf 27}})$,
whereas in the $SO(10)$ model
it produces $3\times({\bf 10}+{\overline{\bf 10}})$.
Therefore, the Calabi--Yau manifold that corresponds to the
${\bf Z}_2\times {\bf Z}_2$ orbifold at the free fermionic point
in the Narain moduli space has $(h_{21}, h_{11})=(27,3)$.

To note the overlap with the construction of ref. \cite{dlow} 
we construct the ${\bf Z}_2\times {\bf Z}_2$ at a generic point in the
moduli space. For this purpose, let us start with the compactified
torus $T^2_1\times T^2_2\times T^2_3$  parameterized by
three complex coordinates $z_1$, $z_2$ and $z_3$,
with the identification
\begin{equation}
z_i=z_i + 1~~~~~~~~~~;~~~~~~~~~~z_i=z_i+\tau_i,
\label{eq:t2cube}
\end{equation}
where $\tau$ is the complex parameter of each torus
$T^2$.
We consider ${\bf Z}_2$ twists and possible shifts of order
two:
\begin{equation}
z_i~\rightarrow~(-1)^{\epsilon_i}z_i+{1\over 2}\delta_i,
\label{eq:z2twistanddance}
\end{equation}
subject to the condition that $\Pi_i(-1)^{\epsilon_i}=1$.
This condition insures that the holomorphic three--form
$\omega=dz_1\wedge dz_2\wedge dz_3$ is invariant under the ${\bf Z}_2$ twist.
Under the identification $z_i\rightarrow-z_i$, a single torus
has four fixed points at
\begin{equation}
z_i=\{0,1/2,\tau/2,(1+\tau)/2\}.
\label{eq:fixedtau}
\end{equation}
The first model that we consider is produced
by the two ${\bf Z}_2$ twists
\begin{eqnarray}
&& \alpha:(z_1,z_2,z_3)\rightarrow(-z_1,-z_2,~~z_3)\cr
&&  \beta:(z_1,z_2,z_3)\rightarrow(~~z_1,-z_2,-z_3).
\label{eq:alphabeta}
\end{eqnarray}
There are three twisted sectors in this model, $\alpha$,
$\beta$ and $\alpha\beta=\alpha\cdot\beta$, each producing
16 fixed tori, for a total of 48. The untwisted sector
adds three additional K\"ahler and complex deformation
parameters producing in total a manifold with $(h_{21}, h_{11})=(51,3)$.
We refer to this model as $X_1$.
This manifold admits an elliptic fibration over a base $F_0={\bf CP}^1\times
{\bf CP}^1$. This can be seen from the  Borcea--Voisin classification
of elliptically fibered Calabi--Yau manifolds \cite{borceaviosin}
and from ref. \cite{mukhi}. 

Next we consider the model generated by the ${\bf Z}_2\times {\bf Z}_2$
twists in (\ref{eq:alphabeta}), with the additional shift
\begin{equation}
\gamma:(z_1,z_2,z_3)\rightarrow(z_1+{1\over2},z_2+{1\over2},z_3+{1\over2}).
\label{eq:gammashift}
\end{equation}
This model has fixed tori from the three
twisted sectors $\alpha$, $\beta$ and $\alpha\beta$.
The product of the $\gamma$ shift in (\ref{eq:gammashift})
with any of the twisted sectors does not produce any additional
fixed tori. Therefore, this shift acts freely.
Under the action of the $\gamma$ shift, half
the fixed tori from each twisted sector are paired.
Therefore, the action of this shift is to reduce
the total number of fixed tori from the twisted sectors
by a factor of $1/2$. Consequently, in this model
$(h_{21}, h_{11})=(27,3)$. This model therefore
reproduces the data of the ${\bf Z}_2\times {\bf Z}_2$ orbifold
at the free-fermion point in the Narain moduli space.
We refer to this model as $X_2$.

The manifold  $X_1$ therefore coincides with the  manifold $X$
of ref. \cite{dlow}, the  manifold  $X_2$ coincides with the
 manifold $Z$, and the $\gamma$--shift in eq. (\ref{eq:gammashift})
coincides with the freely acting involution $\tau_X$ of ref. \cite{dlow}.
Thus, the free
fermionic models admit precisely the structure of the
Calabi-Yau manifolds considered in ref. \cite{dlow}.

\section{Conclusions}

We discussed in this paper the construction of nonperturbative 
flipped $SU(5)$ vacua in Ho\v rava--Witten theory. The flipped $SU(5)$ 
model \cite{sufive} played a pivotal role in the development of the
realistic free fermionic heterotic string models \cite{review}.
Ho\v rava--Witten theory provides the framework to extend the study of 
these models to the nonperturbative domain. Details of these investigations
will be presented in forthcoming publications.

\section*{Acknowledgments}
Support from DGICYT (grant BFM2002-03681) and 
PPARC (grant PPA/A/S/1998/00179 and grant PPA/G/O/2000/00469)
is acknowledged.

\end{document}